\documentclass[aps,prb,reprint,superscriptaddress]{revtex4-1}

\usepackage{amsmath}
\usepackage{amsfonts}
\usepackage{graphicx}
\usepackage{url}

\usepackage{dcolumn}
\usepackage{bm}

\begin{document}

\title{Effect of strain, thickness, and local surface environment on electron transport properties of oxygen-terminated copper thin films}

\author{Alfonso Sanchez-Soares}
\affiliation{Tyndall National Institute, University College Cork, Lee Maltings, Dyke Parade, Cork, T12 R5CP, Ireland}
\author{Sarah L.~T.~Jones}
\affiliation{Tyndall National Institute, University College Cork, Lee Maltings, Dyke Parade, Cork, T12 R5CP, Ireland}
\author{John J.~Plombon}
\affiliation{Intel Corporation, Hillsboro, Oregon 97124, USA}
\author{Ananth P.~Kaushik}
\affiliation{Intel Corporation, Hillsboro, Oregon 97124, USA}
\author{Roger E. Nagle}
\affiliation{Intel Ireland, Collinstown, Leixlip, Co. Kildare, W23 CX68, Ireland} 
\author{James S.~Clarke}
\affiliation{Intel Corporation, Hillsboro, Oregon 97124, USA}
\author{James C.~Greer}%
 \email{jim.greer@tyndall.ie}
\affiliation{Tyndall National Institute, University College Cork, Lee Maltings, Dyke Parade, Cork, T12 R5CP, Ireland}

\begin{abstract}
Electron transport is studied in surface oxidised single-crystal copper thin films with a thickness of up to 5.6 nm by applying density functional theory and density functional tight binding methods to determine electron transport properties within the ballistic regime. The variation of the electron transmission as a function of film thickness as well as the different contributions to the overall electron transmission as a function of depth into the the films is examined. Transmission at the oxidised copper film surfaces is found to be universally low. Films with thickness greater than 2.7 nm exhibit a similar behaviour in local transmission per unit area with depth from the film surface; transmission per unit area initially increases rapidly and then plateaus at a depth of approximately 0.35-0.5 nm away from the surface, dependent on surface facet. Unstrained films tend to exhibit a higher transmission per unit area than corresponding films under tensile strain. 
\end{abstract}

\maketitle

\section{Introduction}

Understanding the effects of scaling to extreme nanoscale limits of a few nanometers on material properties is required for the design of improved nanoelectronic interconnects and devices. Copper has been chosen as the dominant interconnect material in micro- and nanoelectronics since the decision to replace aluminium for integrated circuit applications that occurred approximately two decades ago. As  the size of metallic nanostructures becomes less than the electron mean free path ($ \approx 40$ nm in Cu), significantly more electron collisions with surfaces occur as electrons travel through the sample, which results in a dramatic increase in resistivity. This undesirable increase in line resistance in the nanoscale regime compared to bulk is well established for copper nanostructures~\cite{Steinhogl2002,Zhang2007}. Whether copper can function effectively as an interconnect material in future nanoscale devices remains a subject for debate. The role different structural scattering sources such as defects, surfaces, and grain boundaries~\cite{Fuchs1938,Mayadas1970,Josell2009} play in determining the overall resistance for specific geometries needs to be understood in order to minimise the line resistance of copper and other proposed materials for nanoelectronics interconnects.

The conductivity of metal thin films is extremely sensitive to processing conditions. In the case of gold thin films, Henriquez \textit{et al.}\cite{Henriquez2010,Henriquez2013} report that electron scattering is governed to a large extent by the grain sizes of the film. When the grain size is much smaller than the electron mean free path grain boundary scattering becomes critical to resistivity, conversely surface scattering dominates when the grain size is much larger than the electron mean free path. Grain boundary scattering is reported as the most important contributor to scattering in copper thin films with thickness as low as 27 nm~\cite{Sun2009,Sun2010}. The surface environment itself is also known to significantly affect the conductivity of copper thin films.  Electron scattering at the copper-vacuum surface is partially specular, however variations in the surface environment  can induce dramatic changes to the conductivity of copper thin films. Scattering at the thin film surface is reported to become diffuse after deposition of Ta \cite{Chawla2009,Chawla2011} or surface oxidation\cite{Chawla2010}. 
\textit{Ab initio} simulations suggest that scattering at a copper-metal interface is intimately related to the electronic properties of the metals\cite{Zahid2010}. Resistivity decreases compared to a bare copper surface when the density of states (DOS) of the coating or barrier layer matches that of copper surface atoms (Al and Pd). The converse is also the case with increased resistivity observed for coating metals whose DOS doesn't match copper surface atoms (such as for Ta, Ti, and Ru). However, this resistivity increase on deposition is reported to be reversed by exposure to air, i.e., oxidation of the coating metal for the case of Ta\cite{Rossnagel2004}. Zinc coated copper thin films also show reduced surface scattering compared to bare copper surfaces exposed to air \cite{Zheng2014}. 

In the following, a theoretical study of electron transport properties in oxidised single-crystal copper thin films using \textit{ab initio} and semi-empirical computational techniques is presented. This study explores the effects of different oxidised surface environments, film thickness, and crystallographic orientation along which electronic transport occurs. Consistent with expectations, the transmission at the film surfaces is found to be strongly modulated by the presence of oxygen, and a linear relationship between cross section and transmission for copper films as thin as 2.7 nm is found. 

\section{Method}
\label{ss:method}
\subsection{Thin Film Structures}

\begin{figure} 
\includegraphics[scale=1.0]{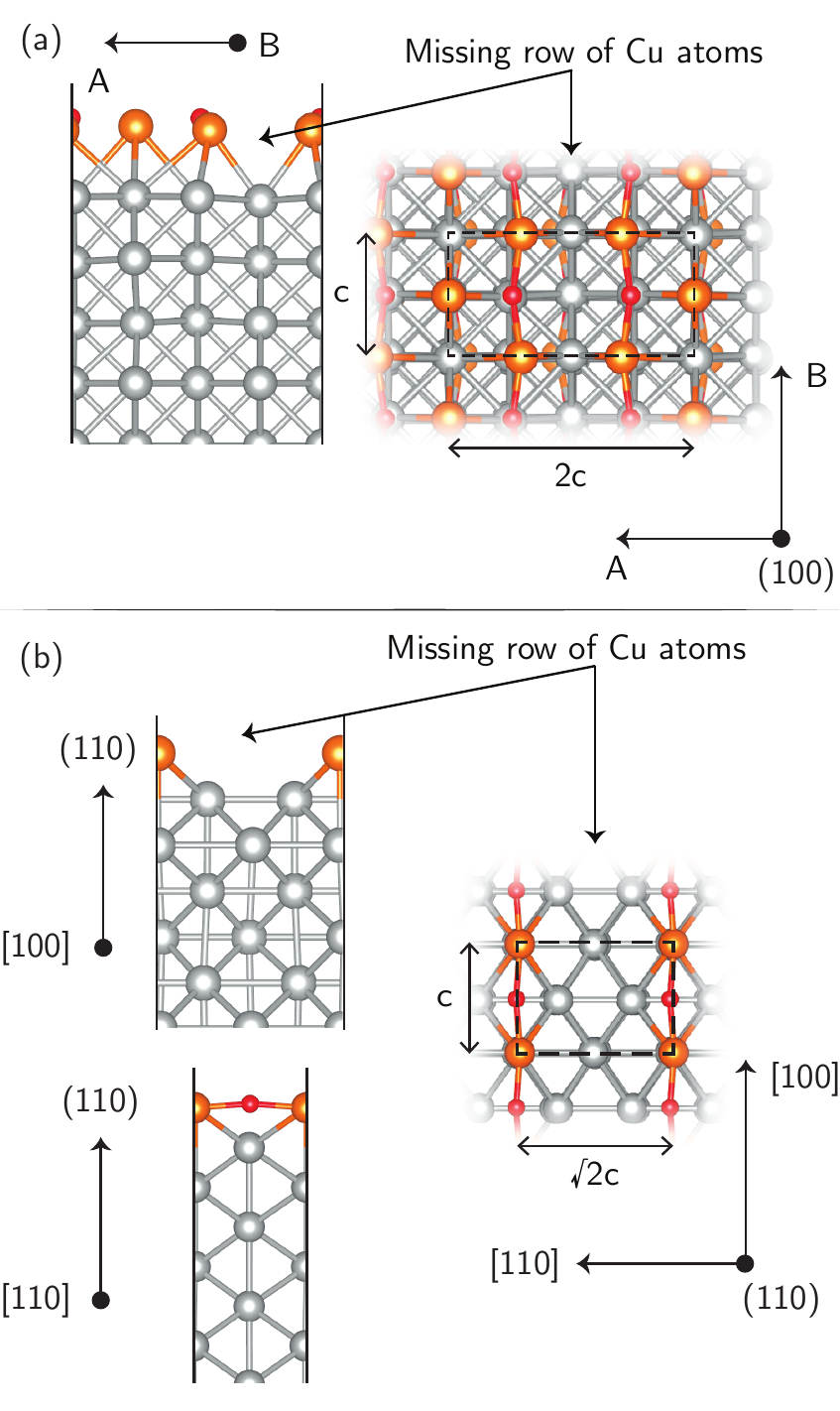} 
\caption{Side and top view of optimised O-terminated Cu thin films. (a) Cu(100)-$( 2 \sqrt{2} \times \sqrt{2} ) R45 ^\circ $-O; and (b) Cu(110)-$p( 2 \times 1) $-O. Copper atoms away from surfaces are shown in silver.}
\label{fig:Structure}
\end{figure}

The thin films in this study are based on bulk fcc Cu with two different exposed facets on the surface. The first set of structures have exposed (100) facets at their surfaces in which there is a monolayer of oxidised copper with an oxygen coverage of 0.5 ML, and present a local surface environment with a $( 2 \sqrt{2} \times \sqrt{2} ) R45 ^\circ $ missing row reconstruction [see Fig. \ref{fig:Structure}(a)]; a structure which has previously been theoretically and experimentally identified \cite{ZENG1989,ATREI1990,Kittel2001,Bonini2006,Harrison2006,Duan2010,Baykara2013} . The immediate surface of the simulated cell consists of 3 Cu atoms and 2 O atoms; each O atom bonds to 4 Cu atoms, three at the surface and one in the Cu layer below the surface, while each surface Cu atom bonds to 2 O atoms. A two dimensional oxide forms with a missing row of Cu atoms at the surface [as indicated in Fig.~\ref{fig:Structure}] with the O atoms inserting into Cu-Cu bonds adjacent to the missing row. This structural rearrangement enables the O-Cu-O angle to be close to 180$^{\circ}$ as occurs in bulk copper oxides such as CuO and Cu$_2$O. 

Films with thicknesses ranging from 1.9 to 5.6 nm are considered; and in addition the effect of tensile strain on conductance is investigated by studying structures generated with an increased lattice parameter $c$ from 3.63 \AA{} --the equilibrium lattice parameter within our approximation-- to 3.80 \AA{} and 3.90 \AA{} (biaxial strains of 4.7\% and 7.4\%, respectively). Interconnect copper structures are expected to present regions with localised strain near interfaces with diffusion barrier layers, which tend to have greater lattice spacings than copper. As dimensions reduce and due to the malleability of metals, the portion of a copper line impacted by strain becomes increasingly important. In the case of ultra-thin Cu nanowires, previous work reports lattice parameter tendency to increase due to the presence of surface oxidation with Cu-Cu spacings increasing as much as 11\%\cite{PhysRevB.92.115413}. We therefore include the case of tensile strained films in order to assess its impact on the conductivity of copper nanostructures.

Transport along [110] and two perpendicular non-equivalent [100] directions are considered for this set of structures: along axes A and B [see Fig. \ref{fig:Structure}(a)]. Whilst away from the surface both directions are equivalent to [100] in fcc bulk Cu, the surface environment introduces anisotropy: the missing row of Cu atoms is perpendicular to axis A and parallel to axis B. 

In order to further explore the effect of different surface environments, a second set of structures with (110) exposed facets, an oxygen coverage of 0.5 ML, and thicknesses ranging from 2.3 to 5.1 nm is studied. In these structures, the surface environment corresponds to the missing row $p(2\times1) $ reconstruction which has been extensively studied \cite{BESENBACHER1993,Uehara2002,Kishimoto2008,Duan2010,JunLi2016}, and has been the focus of recent atomic force microscopy studies due to its chemical fingerprinting properties \cite{Bamidele2012,Bamidele2014,Bamidele2014b}. Similar to the previously discussed surface environment, each surface Cu atom bonds to 2 O atoms, while each O atom bonds to 4 Cu atoms: two at the surface and two in the next Cu layer into the film. In these structures, electron transport properties are studied along non-equivalent perpendicular directions which deep into the film correspond to transport along the [100] and [110] crystallographic directions of bulk Cu, respectively [see Fig. \ref{fig:Structure}(b)].

The definition of film thickness is somewhat ambiguous due to the atomistic structure of the films; a convention of choosing the thickness of the slab to be the internuclear separation between atoms in the top and bottom surfaces of the thin film plus twice the atomic radius of Cu (128 pm) is followed. 

\subsection{Computational Details}

The electronic properties of Cu thin films are studied using density functional theory (DFT) as implemented by the software package OpenMX\cite{OpenMX}; and density functional tight binding (DFTB) as implemented by QuantumWise\cite{QW,Brandbyge2002,Soler2002}. The PBE\cite{PERDEW96} formulation of the generalised gradient approximation (GGA) exchange-correlation functional is used throughout. Norm-conserving pseudopotentials\cite{MORRISON93} and a strictly localised pseudo-atomic orbital (PAO) basis\cite{OZAKI03,OZAKI04} are used. 
The basis sets used are 6.0H-s4p2d2 and 7.0-s2p2 for Cu and O respectively. The first part of the basis set notation gives the PAO cut-off radius in Bohr, while the second part indicates the orbitals used for the valence electrons, e.g., O 7.0-s2p2 corresponds to a cut-off radius of 7.0 Bohr and 8 basis functions (2 s functions and 6 p functions). Structural parameters derived from this Cu basis set agree well with experiment, giving an optimal fcc lattice parameter of 3.63 \AA\ and a bulk modulus of 135 GPa compared to experimental values of 3.615 \AA\ and 137 GPa, respectively. A supercell approach is taken whereby the thin films are modelled with periodic boundary conditions in the plane of the film and a minimum of 1 nm of vacuum is introduced in the direction normal to the thin films' surface to prevent interaction of periodic images. The Brillouin zones of the two sets of structures are sampled using a $11 \times 5 \times 1$ and $ 7 \times 9 \times 1$ k-point grid generated according to the Monkhorst-Pack method\cite{Monkhorst1976}, respectively; and a grid corresponding to an energy cut-off of 200 Ry is used for numeric integration of real-space quantities.

The upper and lower surfaces of the thin films are optimised separately with a frozen bulk Cu back plane as shown in Fig.~\ref{fig:Structure} until forces acting on individual atoms remain below $3\times10^{-4}$ Hartree/Bohr.
After optimisation of the local surface environment containing the oxide, separate upper and lower surfaces about 0.8 nm thick are joined together and the thickness of the resulting film is increased by addition of bulk-like fcc Cu, with suitable lattice parameter, at the center as required to achieve the desired thickness.

The electronic transport properties of the resulting structures are studied in the ballistic transport regime using an approach based on Green's functions\cite{Brandbyge2002} within the context of the Landauer-B\"{u}ttiker formalism\cite{Landauer1957,Datta1995}. 

The maximum current that can flow through a finite central region connected to two semi-infinite leads is related to the probability of an electron being transmitted through it via the relation 

\begin{eqnarray}
\label{current_eq}
I(V) = \frac{e}{h} \sum_{\sigma} \int T_{\sigma} (E,V)& \nonumber \\
\times \left[ f  \left( E , \mu_R , \right. \right. & \left. \left. T_R \right) -
f \left( E, \mu_L , T_L \right) \right] \, dE \, ,
\end{eqnarray}
where $e$ is the electron charge, $h$ is Planck's constant, $T_{\sigma} (E,V)$ is
the transmission coefficient per spin channel $\sigma$ at energy $E$ and applied bias $V$, 
$f$ is the Fermi-Dirac distribution, $T_{L}$ ($T_{R}$) and $\mu_L$ 
($\mu_R$) are the temperature and chemical potential  
of the left (right) electrode, and the applied bias is given by

\begin{equation}
V=\frac{\mu_R - \mu_L}{e}.
\end{equation}

For small applied biases we can approximate the conductance by its linear-response zero-temperature limit in the non-spin polarised case as

\begin{equation}
\label{eq:LR}
G =  \frac{2e^2}{h} T (E_F,V=0).
\end{equation}

At finite temperatures, the largest contributions to the integral in Eq. \ref{current_eq} will come from the transport of electrons with energies closer to the Fermi level; in particular, assuming slow variations of the transmission coefficients with energy, the previous result can be extended beyond the zero-temperature limit\cite{Datta1995}.

Throughout this work, we have computed the transmission coefficients over a 100 meV energy range centered about the Fermi level $T(E_F \pm 50 ~meV,V=0)$ as a measure of the films' maximum conductance within the linear-response approximation. This allows an assessment of the variation of transmission coefficients around the Fermi level -- and hence the validity of Eq. \ref{eq:LR} -- and a more robust benchmark of different methods for computing transport properties.

Integration of quantities in electronic transport calculations is performed by sampling with 301 k-points along the cell direction parallel to the transmission direction and a single k-point along the non-periodic direction. The computational demand involved in the calculation of transmission coefficients using first principles DFT becomes prohibitively large as the thickness of the films increase. 
Therefore we assess the suitability of semi-empirical DFTB for calculating electron transmission probabilities in copper nanostructures with oxidised surfaces, and under tensile strain. 
Two sets of DFTB parameters are employed and compared against first principles DFT calculations. 
The first set of parameters is the matsci set (version 0-3)~\cite{www.dftb.org,NJardillier2006} and a second parameter set labelled 'custom', which has been optimised specifically to have a physical band gap for bulk Cu$_2$O, which is not reproduced from the matsci parameters.

In the DFTB method\cite{Seifert1996,Seifert2012}, two center interactions are computed by using Slater-Koster integral tables which describe the Hamiltonian and overlap matrix elements between atoms on a equidistant grid\cite{Slater1954}. The 'custom' DFTB parameter set was generated by starting from all-electron DFT wave functions and optimising interactions between pairs of atoms such that when put into a bulk crystal structure, the experimental band structure of the crystal was recovered. These differ from the matsci parameter set in that specific orbitals and wave functions were targeted to be optimised in order to fit experimental band structures. The procedure followed in their generation is similar to that described by \citet{Wahiduzzaman2013}.

These DFTB calculations also serve as the basis for the determination of transmission pathways\cite{Solomon2010} within the thin films, which provide a spatially resolved projection of total electron transmission into local coefficients defined between pairs of atoms: For any plane perpendicular to the transmission direction which divides the system in two regions A and B, it holds that
\begin{equation}
T(E) = \sum_{i \in A, j \in B} T_{ij}(E), \\
\label{eq:pathways}
\end{equation}
where $i$ and $j$ denote atoms located on each side of said plane.
In the case of perfectly periodic systems such as those studied here, the use of a Green's function approach is not necessary in order to obtain the total transmission coefficients at zero applied bias $T(E,V=0)$, but it is required for computing their projection into spatially localised transmission coefficients.

\section{Results and discussion}

\begin{figure}
\includegraphics[scale=1]{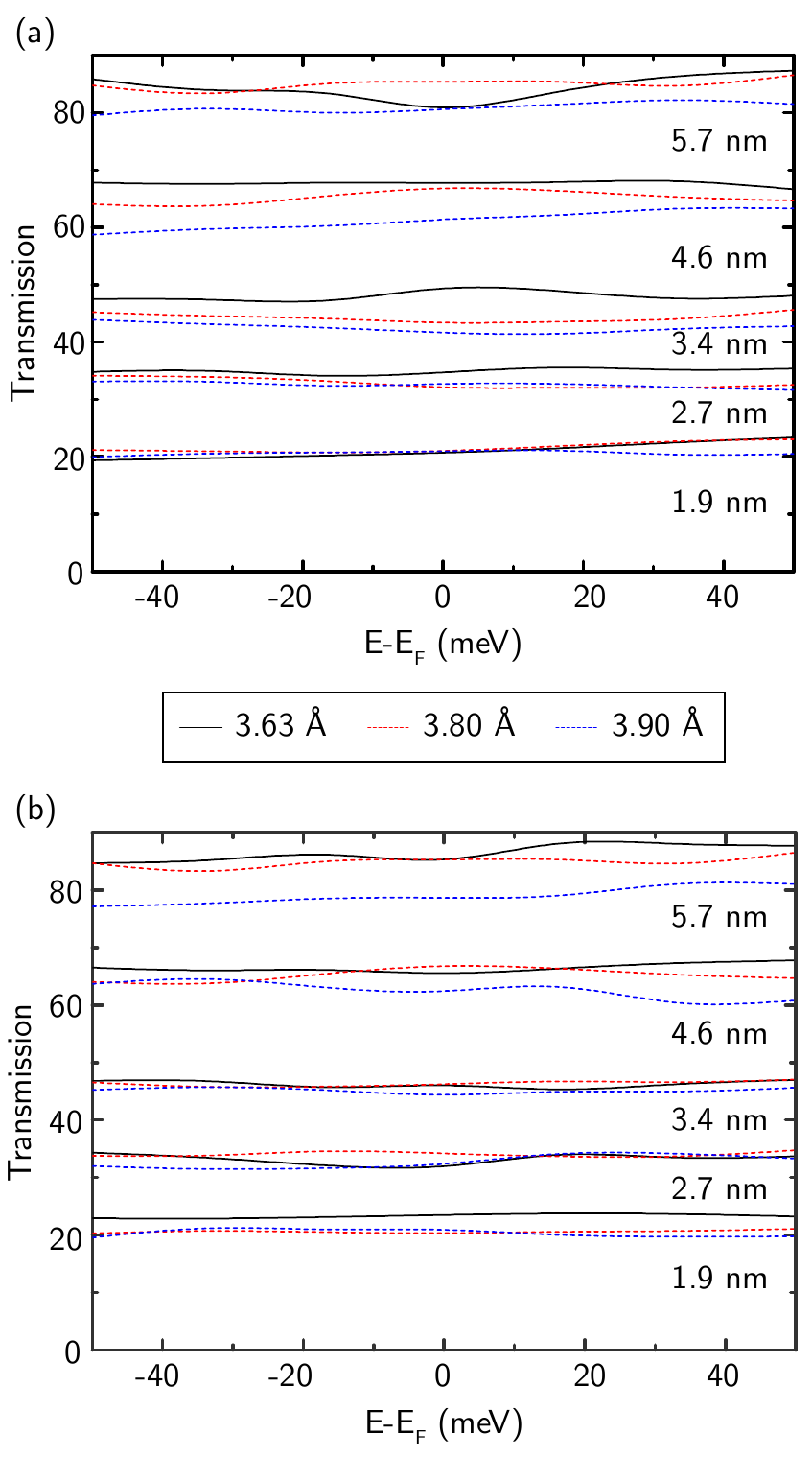}
\caption{DFT-computed transmission for film thickness of 2.0, 2.7, 3.4, 4.5 and 5.6 nm with varying values of the lattice parameter -$c$- for inequivalent transport directions (a) A and (b) B as indicated in Fig. \ref{fig:Structure}(a). Zero of energy is taken to be at the Fermi level.}
\label{fig:Cell-Tran}
\end{figure}

\subsection{DFT vs. DFTB: charge transport properties}
\label{ss:dftb}
The DFT calculated total transmission as a function of film thickness and cell parameter $c$ for directions A and B is shown in Fig.~\ref{fig:Cell-Tran}.
Inspection of the transmission per unit area at the Fermi level reveals a decrease in the maximum current density that can flow through the films with increasing tensile strain [see Table \ref{tab:strain}]: As the metal atoms in the structure become further apart and the overlap between their orbitals becomes smaller, the number of conducting paths decreases as it becomes more difficult for electrons to propagate between atoms.

\begin{figure} 
\includegraphics[scale=1]{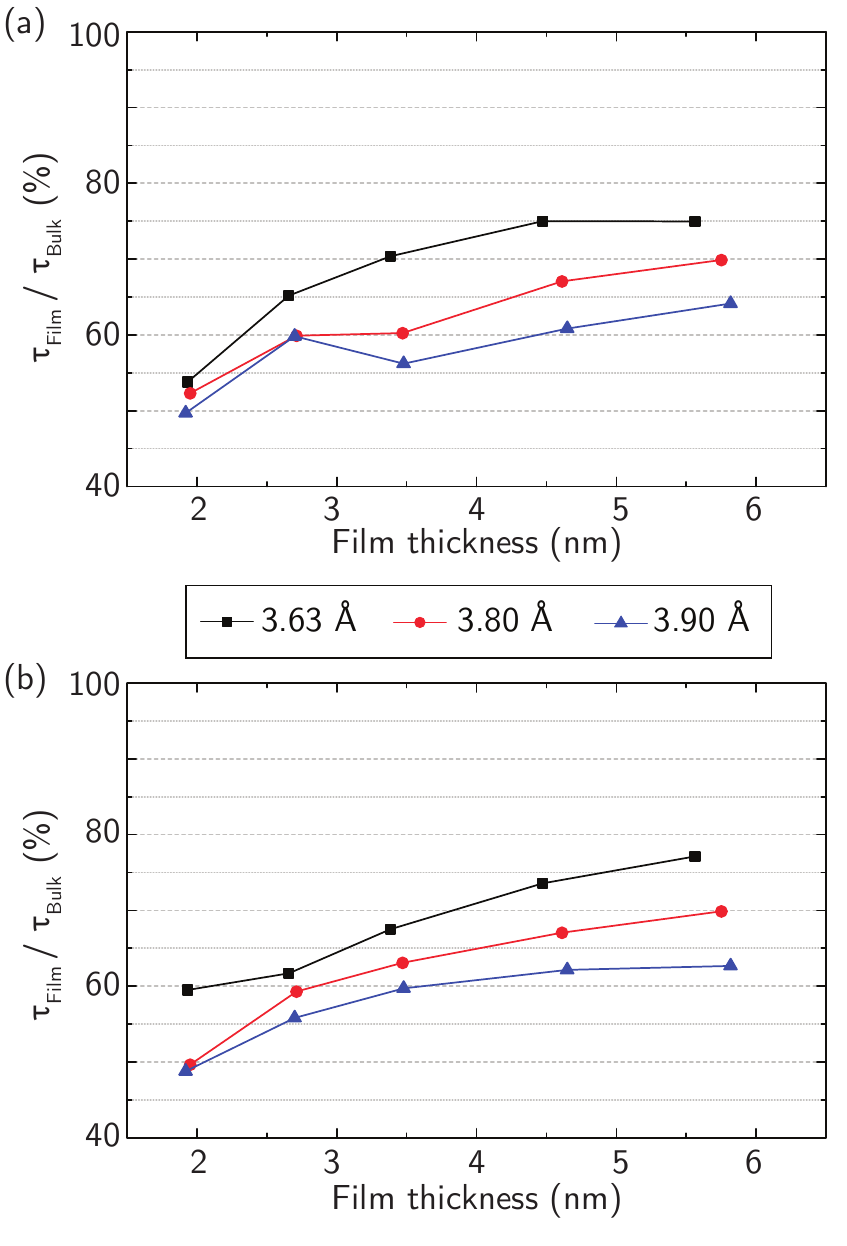}
\caption{DFT calculated ratios between the transmission per unit area of structures shown in Fig. \ref{fig:Structure}(a) and bulk Cu. The lengths labelling each curve in the legend indicate the lattice constant (biaxial strain).}
\label{fig:Strain}
\end{figure}

Table \ref{tab:strain} also shows an increase of transmission per unit area at the Fermi level $\tau(E_F)$ with increasing thickness: The addition of new conducting paths further away from the surface, and hence with a lower associated probability to scatter off it, results in a net increase of the total current density that flows through the thicker films. Figure \ref{fig:Strain} illustrates these findings by plotting the ratio between the transmission per unit area of the films and that of bulk copper versus film thickness for all considered strains, which is equivalent to the ratio between their conductances.

\begin{table}[t]
\caption{\label{tab:strain}DFT calculated transmission per unit area at $E_F$ for structures shown in Fig. \ref{fig:Structure}(a) with varying thickness and biaxial tensile strain.}
\begin{ruledtabular}
\begin{tabular}{l *{3}{c}}
 &  \multicolumn{3}{c}{$\tau(E_F)$ (nm\textsuperscript{-2})}\\
Strain $\to$ & 0 \% & 4.7 \% & 7.4\%  \\
\hline
\\
(100)-{[}100{]} A   \\
~~~~1.9 nm & 7.34 & 7.08 & 7.00 \\
~~~~2.7 nm & 8.99 & 8.15 & 8.35 \\
~~~~3.4 nm & 10.04 & 8.22 & 7.67 \\
~~~~4.5 nm & 10.43 & 9.52 & 8.46 \\
~~~~5.6 nm & 10.02 & 9.76 & 8.88 \\
(100)-{[}100{]} B  \\
~~~~1.9 nm & 8.32 & 6.81 & 6.95 \\
~~~~2.7 nm & 8.26 & 8.27 & 7.67 \\
~~~~3.4 nm & 9.35 & 8.75 & 8.17 \\
~~~~4.5 nm & 10.09 & 9.52 & 8.60 \\
~~~~5.6 nm & 10.57 & 9.76 & 8.67 \\
\end{tabular}
\end{ruledtabular}
\end{table}

\begin{figure}
\includegraphics[scale=1]{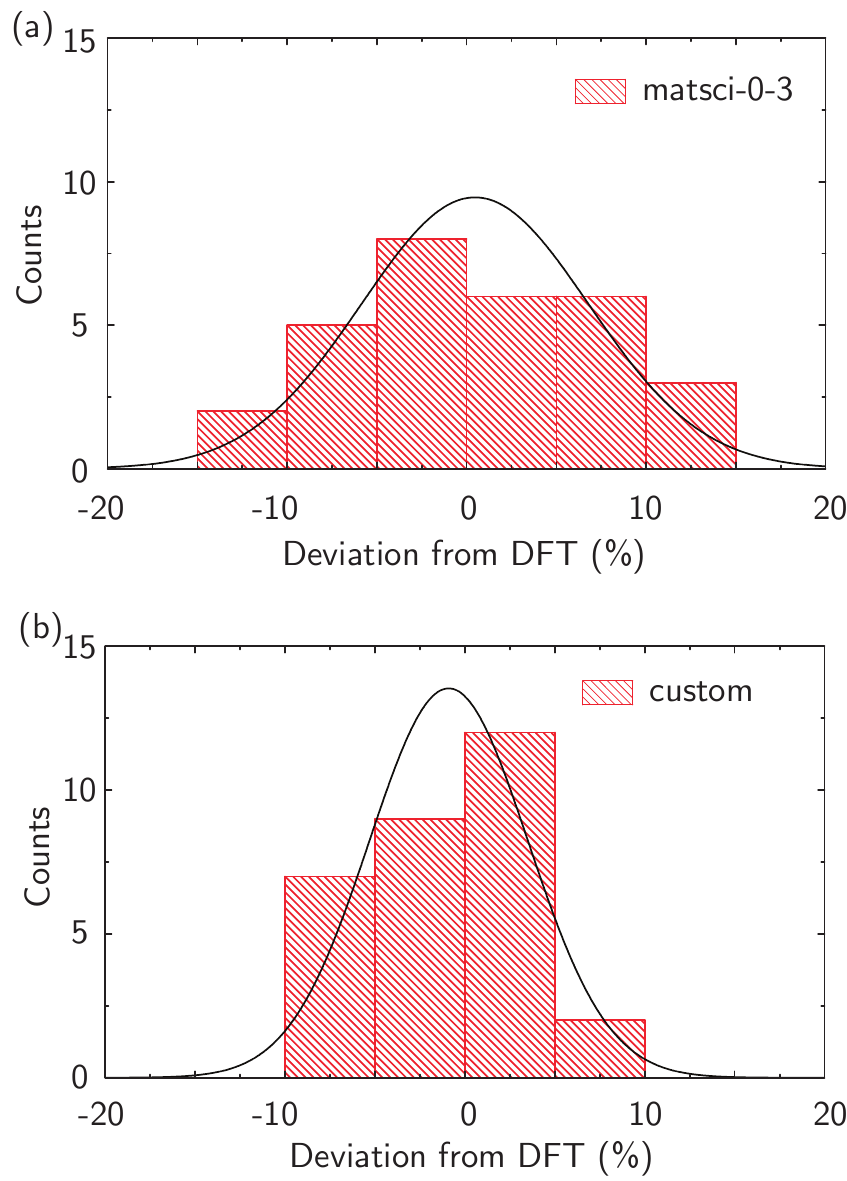}
\caption{Histograms showing the error in transmission coefficients obtained using DFTB with respect to DFT transmission at $E_F$ obtained with (a) matsci, and (b) custom parameter sets. Values for structure set shown in Fig. \ref{fig:Structure}(a) with varying thickness and strain have been compared.}
\label{fig:Histograms}
\end{figure}

\begin{figure}
\includegraphics[scale=1.0]{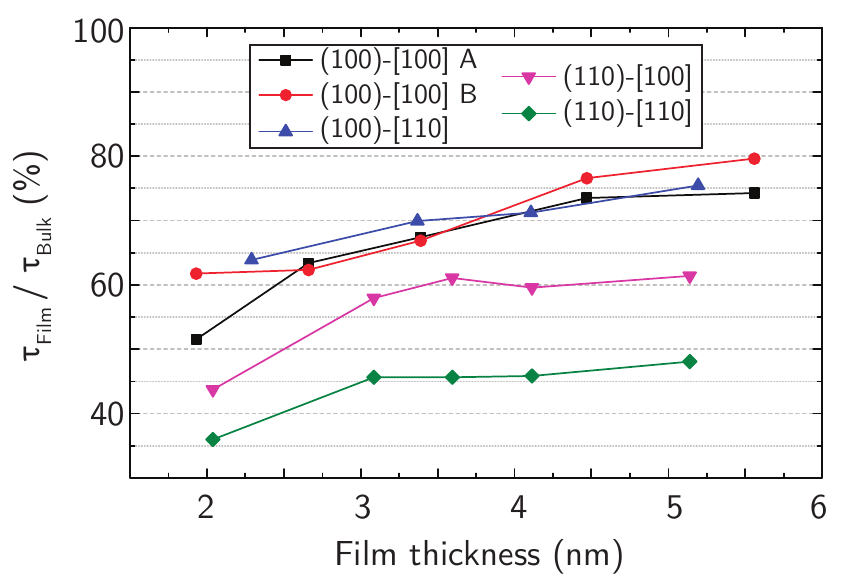}
\caption{Evolution of calculated transmission per unit area with film thickness for both structure sets shown in Fig. \ref{fig:Structure} along the [100] and [110] transport directions. DFTB with the custom parameter set has been employed.}
\label{fig:Tran-Area}
\end{figure}

Similar behaviour is observed in DFTB calculations using both parameter sets. Transmission coefficients for all cases shown in Fig. \ref{fig:Cell-Tran} have been computed using both parameter sets. Histograms showing deviations from DFT transmission values at the Fermi level are shown in Fig. \ref{fig:Histograms}. The custom parameter set is found to provide a description of the films' electron transport properties closer to that of DFT, with a maximum error of 10\% and a standard deviation of 4.4 versus values found for the matsci parameter set of 13\% and 6.3, respectively.
DFTB hence provides a reasonable approximation for transmission in Cu/O systems at a fraction of the computational cost of DFT and is able to describe thin films under tensile strain with similar accuracy. Figure \ref{fig:Tran-Area} illustrates the evolution of film conductance with thickness depending on transport direction and local surface environment by showing the transmission per unit area of both structure sets and transport directions with respect to that of bulk copper for increasing film thickness, as computed with DFTB using the custom parameter set.

\subsection{Transmission pathways}

\begin{figure}
\includegraphics[scale=1]{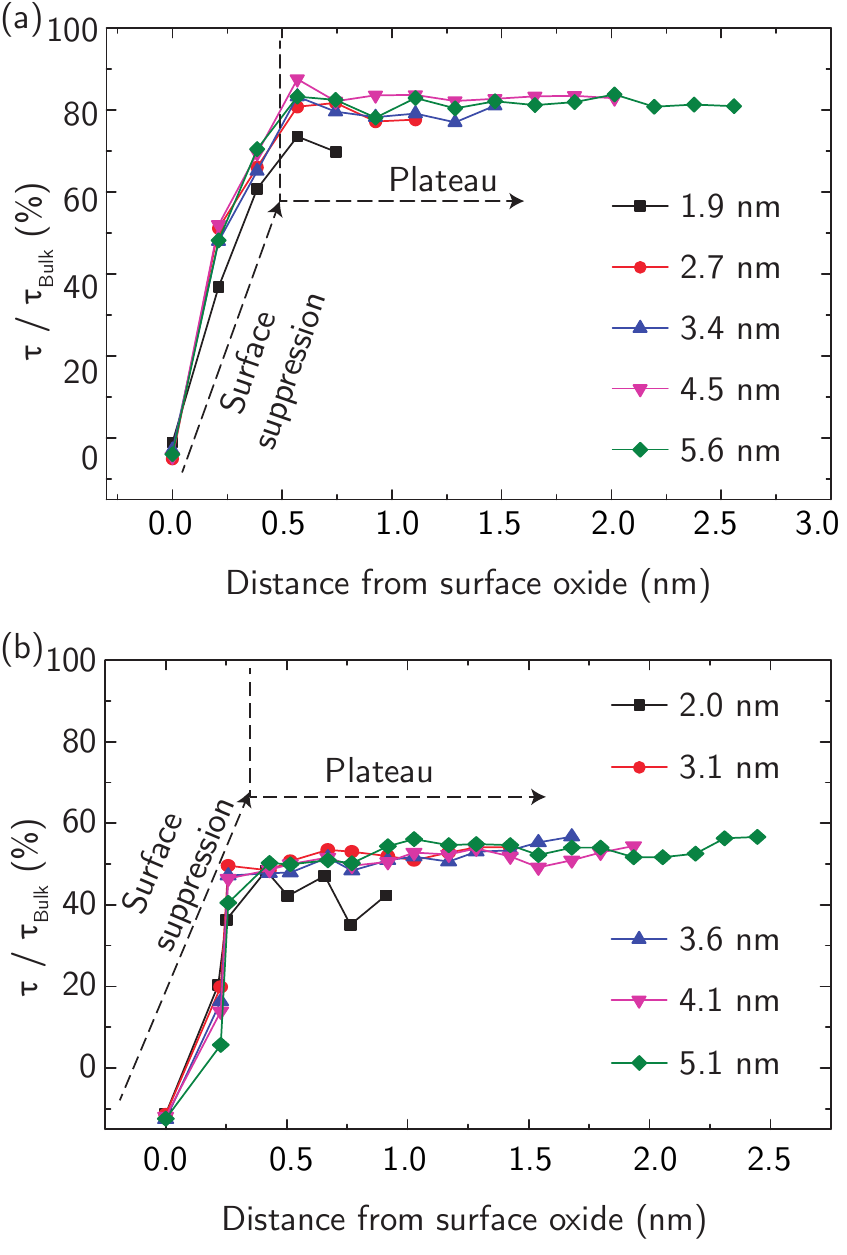}
\caption{Transmission per unit area at the Fermi level as a function of distance from the thin film's surface oxide. Transmission is strongly suppressed near films' surfaces, below the surface transmission per unit area initially increases rapidly and plateaus for depths greater than $\approx$ 0.35-0.5 nm, dependent on surface environment.}
\label{fig:Slab-depth}
\end{figure}

The total transmission coefficients provide information about the total number of electron transmission channels through the films as a whole, whilst the use of transmission pathways allows for a spatially resolved analysis of charge transport. Transmission pathways calculations have been performed using DFTB with the custom parameter set for both transport orientations of each structure set at the Fermi level $E_F$.  
These localised pathways between individual atom pairs within the thin films enable an extraction of local electron transmission coefficients as a function of depth into the copper films, as shown in Fig. \ref{fig:Slab-depth}. 
The calculated transmission per unit area near the oxidised surfaces is much lower than at the center of the films, which reveals that the reduction of propagating states caused by the presence of the oxide and its modification of the local electronic structure extends into the film and is not limited to the surface layers.
Below the surface oxide layer, transmission per unit area initially increases rapidly with depth and plateaus for distances greater than $\approx$ 0.5 nm away from the surface in structures with $(100)$ surface facets [Fig. \ref{fig:Structure}(a)]; while plateau values are observed for depths greater than $\approx$ 0.35 nm in structures with $(110)$ surface facets [Fig. \ref{fig:Structure}(b)]. This variation in depth up to which the presence of the surface oxide suppresses electron transport is attributed to differences in the local surface environment with varying crystallographic orientations. Values of the transmission per unit area for the thinnest film studied in each structure set (1.9-2.0 nm thick) remain below the maxima seen for the thicker films.  With the exception of said films, the depth dependence of the transmission per unit area is remarkably uniform for a given structure set and transport direction. A maximum value of transmission per unit area of 50-80\% that of bulk copper is reached for depths greater than 0.35-0.5 nm, depending on surface environment and transport direction, in all studied structures thicker than 2 nm, remaining approximately constant all the way through to the center [see Fig. \ref{fig:Slab-depth}]. 
However, these plateau values remain below the full transmission per unit area of bulk copper despite their apparent convergence at the center of the films, an effect brought on by the presence of surface scattering and its limiting of the number of conduction paths for electrons to propagate along the film.

While the center of the films exhibit an atomic structure similar to that of bulk Cu, and hence one may expect the local number of conducting paths to be the same, the close proximity of surfaces imposes restrictions over the number of independent paths in the films, even at the center. The reason behind these differences might be found in the electronic structure of bulk copper: although it exhibits an isotropic conductivity, inspection of the state distribution across its Brillouin zone (BZ) reveals that not all directions are equivalent in terms of states which propagate along them at the Fermi level: whilst states can be found along $\Sigma$ and $\Delta$ paths --associated with $[110]$ and $[100]$ directions, respectively--, there are no states along the $\Lambda$ path --associated with direction $[111]$--; a feature made evident by the shape of the Fermi surface and its vanishing DOS along directions perpendicular to the hexagonal faces of the BZ\cite{Segall1962}.

It is worth noting that the depth of influence of the surface on electronic transport properties found in this study may be modified when introducing the effects of inelastic sources of scattering and decoherence effects into the model.

\begin{figure} 
\begin{center}
\includegraphics[scale=1]{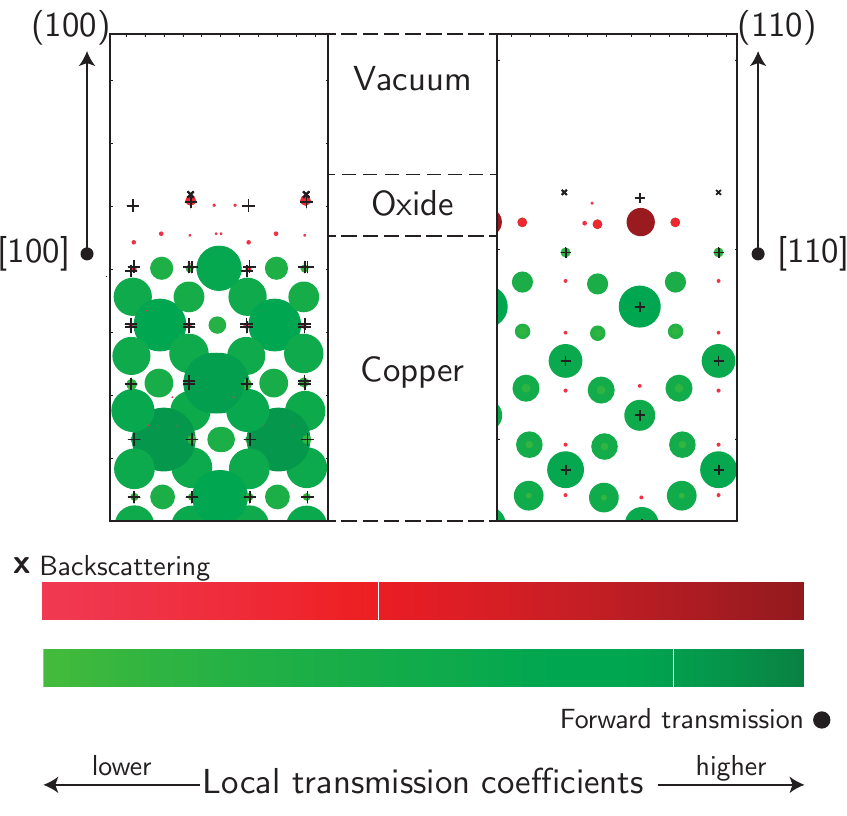}
\caption{Projections of transmission pathways at the Fermi level at a plane perpendicular to the transport direction and located halfway between atomic planes. Both the size of the bubbles and their colour represent the magnitude of each pathway. Cu atoms are represented by \textbf{+} signs, while O atoms at the surface are shown as \textbf{x}. Transport along direction A of the structure shown in Fig. \ref{fig:Structure}(a) is shown on the left; while the plot on the right corresponds to transport along [110] in the structure shown in Fig. \ref{fig:Structure}(b).}
\label{fig:projections}
\end{center}
\end{figure}

\subsection{Surface scattering model}

The previous analysis of the transmission pathways results suggests that the total transmission in all but the thinnest films of each structure set can be decomposed as a sum of two contributions: a \emph{surface} contribution describing transmission for depths less than 0.35-0.5 nm into the films and a \emph{plateau} contribution describing transport for greater depths. In terms of transmissions per unit area ($\tau$), we may write:

\begin{equation}
\label{eq:greer}
T = A^{Surf}\tau_{Surf} + A^{Plat}\tau_{Plat},
\end{equation}
where we have split the cross-sectional area of the film into the area of the surface and plateau regions, and each term on the rhs is proportional to the maximum current that can flow through the corresponding section of the film.
Since the depth up to which transmission is strongly suppressed by the surface does not vary with thickness, we shall consider $A^{Surf}$ to be constant for all structures -- except the thinnest films -- within each set.
We then proceed to fit Eq. \ref{eq:greer} in order to obtain values for $\tau_{Surf}$ and $\tau_{Plat}$ at the Fermi level.

\begin{table} 
\caption{\label{tab:greer}Computed values for local transmission per unit area with respect to bulk Cu for both structure sets using the custom DFTB parameter set. Values have been obtained by performing a linear fit of obtained data to Eq. \ref{eq:greer}; correlation coefficients greater than 0.99 were obtained in all cases. Values shown in parenthesis have been obtained wit DFT.}
\begin{ruledtabular}
\begin{tabular}{l *{2}{c}}
& $\tau_{Surf}/\tau_{Bulk} (\%)$  & $\tau_{Plat}/\tau_{Bulk} (\%)$ \\
\hline
\\
(100)-{[}100{]} A & 38 (54) & 85 (77)  \\
(100)-{[}100{]} B & 19 (23) & 97 (88) \\
(100)-{[}110{]} & 39 & 86 \\
\\
(110)-{[}100{]} & 44 & 65 \\
(110)-{[}110{]} & 36 & 52 \\
\end{tabular}
\end{ruledtabular}
\end{table}
 
Table \ref{tab:greer} shows the obtained values for transmission per unit area on each region for both structure sets and transport directions, scaled by values obtained for bulk Cu. For the first set of structures, A and B transport directions correspond to bulk [100] deep into the film  as asymmetry is only introduced near the surface. It is seen that whilst the surface environment for electrons propagating along the B axis -- in which the missing rows of Cu atoms are parallel to the transport direction -- is more resistive, it has less of an impact on the resistance of the plateau region than for electrons travelling along the A axis. This correlates to the fact that total film transmission per unit area increases at a higher rate for transport along the B axis as thickness increases, as can be seen in Fig. \ref{fig:Tran-Area}. In the case of these two transport directions, we have included results obtained with DFT as an assessment of the error introduced by the use of DFTB on the values of fitted parameters: whilst differences on $\tau_{Surf}$ can be as high as 40\%, qualitative trends hold and errors on $\tau_{Plat}$ remain within the expected 10\%.

Results for transport along the [100] crystallographic direction for the second set of structures show its corresponding surface environment's resistance to be similar to that of transport along (100)-[100] A, although it results in a higher resistance at the plateau region, a feature evident in the fact that its transmission per unit area increases less than for structures in the first set in the 3-5 nm thickness range [see Fig. \ref{fig:Tran-Area}]. Finally, results for transport along the [110] direction show both surface environments to be similarly resistive, with structures with (110) surface facets exhibiting lower transmission per unit area in the plateau region. Overall, the set of structures with (100) surfaces exhibit stronger anisotropy, as shown by the vertical separation between curves corresponding to both of its transport directions in Fig. \ref{fig:Tran-Area}.

Figure \ref{fig:projections} shows a visual representation of transmission pathways crossing a plane perpendicular to the transport direction, and located halfway between two atomic planes. Two cases are illustrated: (100)-[100] A and (110)-[110]. Transmission along [100] is shown to always take place at an angle with respect to the transport direction; while transport along [110] is shown to have a sizeable parallel component. This is shown by the fact that the largest transmission pathways (bubbles) along [100] are found halfway between Cu atoms (\textbf{+}), whilst the largest bubbles are found to lay directly on top of Cu atoms for transport occurring along [110]. Forward transmission is found to be severely suppressed when approaching both surfaces, with an associated increase in backscattering coefficients.

\section{Conclusions}

It is found that the DFTB calculations of electron transmission performed provide a reasonable approximation to DFT for Cu/O systems, with the custom parameter set performing closer to DFT (error $<10\%$) when compared to the matsci set (error $<13\%$).
This enables the calculation of transport properties of structures with a larger number of atoms and hence more realistic models of nanoscale interconnects.
It is found that oxidation of the surface of copper thin films results in a dramatic reduction of electron transmission directly at the interface, an effect which severely degrades local electron transmission coefficients for distances of up to approximately 0.35-0.5 nm into the films, depending on the crystallographic orientation of exposed surface facets -- a result which highlights the importance and provides a scale length for interface effects in nanoscale copper systems -- although introducing the effects of inelastic scattering and other decoherence effects might modify surface effects' depth of influence discussed in this work. Maximising conductance in nanoscale copper interconnects requires careful interface design between conductor and barrier layer as the chemical environment at the surface and orientation along which transport occurs can have critical impact for Cu systems with feature lengths below 10 nm.
Tensile strain tends to be detrimental to film conductance as the transmission per unit area is found to be reduced for films under biaxial tensile strain studied in this work. For film thicknesses of $\approx$ 3 nm and greater, a core region for which transmission is somewhat independent of depth is formed, although bulk-like transmission has not been observed even at the center of films as thick as 5.6 nm.
The use of a localised transmission model shows the resistance of the studied films to dramatically increase near the surface; whereas the increase in resistance at the core region is found to strongly depend on the surface environment and transport direction.

\begin{acknowledgments}

This work was performed as part of the Intel-Tyndall research collaboration sponsored by Intel Components Research. A.S.S. was funded under an Irish Research Council postgraduate scholarship. Partial funding was provided by the Science Foundation Ireland Investigator program, Grant No. 13/IA/1956. We are grateful to QuantumWise A/S for support and access to the QuantumWise simulation software. Atomistic visualisations were rendered using VESTA \cite{VESTA}. 
\end{acknowledgments}

\end{document}